\documentstyle[prl,aps,amssymb,amsmath,multicol,epsf,psfrag]{revtex}

\begin{document}

\title{Quantum interferometric optical lithography:\\ 
	towards arbitrary two-dimensional patterns}
\author{Pieter Kok\cite{pieter}$^1$, Agedi N.\ Boto$^2$, Daniel S.\ Abrams$^2$,
	Colin P.\ Williams$^2$, Samuel L.\ Braunstein$^1$ \\ and 
	Jonathan P.\ Dowling$^2$}
\address{$^1$Informatics, Bangor University, Bangor LL57 1UT, UK}
\address{$^2$Jet Propulsion Laboratory, California Institute of Technology, \\
	 Mail Stop 126-347, 4800 Oak Grove Drive, Pasadena, California 91109}

\maketitle

\begin{abstract}
 As demonstrated by Boto {\em et al}.\ [Phys.\ Rev.\ Lett.\ {\bf 85}, 2733
 (2000)], quantum lithography offers an increase in resolution below the 
 diffraction limit. Here, we generalize this procedure in order to create 
 patterns in one and two dimensions. This renders quantum lithography a
 potentially useful tool in nanotechnology.
\end{abstract}

\medskip


\begin{multicols}{2}

Optical lithography is a widely used printing method. In this process light is 
used to etch a substrate. The exposed or unexposed areas on the substrate then 
define the pattern. In particular, the micro-chip industry uses lithography to 
produce smaller and smaller processors. However, classical optical lithography 
can only achieve a resolution comparable to the wavelength of the light used  
\cite{brueck98,mack96,mansuripur00}. It therefore minimizes the scale of the 
patterns. To create smaller patterns we need to venture beyond this classical 
boundary \cite{yablonovich99}. In Ref.\ \cite{boto00} we introduced a procedure
called {\em quantum} lithography that offers an increase in resolution beyond 
the diffraction limit. This process allows us to write closely spaced lines in 
one dimension. However, for practical purposes (e.g., optical surface etching) 
we need to create more complicated patterns in both one and two dimensions. 
Here, we study how quantum lithography can be extended to create these
patterns.

This paper is organized as follows: first, for completeness, we present a 
derivation of the Rayleigh diffraction limit. Then, in Sec.\ \ref{intro} we 
reiterate the method introduced in Ref.\ \cite{boto00}. Then, in Sec.\ \ref{1D}
we give a generalized version of the states used in this procedure. We show 
how we can tailor arbitrary one-dimensional patterns with these states. In 
Sec.\ \ref{2D} we show how four-mode entangled states lead to patterns in two 
dimensions. Sec.\ \ref{phys} addresses the physical implementation of quantum 
lithography.

\section{Classical resolution limit}\label{resolution}

When we talk about optical resolution, we can mean two things: it may denote 
the minimum distance between two nearby points which can still be resolved 
with microscopy. Or it can denote the minimum distance separating two points 
which are printed using lithography. In the limit of geometric optics these 
resolutions would be identical. In this section we derive the classical 
resolution limit for interferometric lithography using the so-called Rayleigh 
criterion \cite{rayleigh1879}. 

Suppose two plane waves characterised by $\vec{k}_1$ and $\vec{k}_2$ hit a 
surface under an angle $\theta$ from the normal vector. The wave vectors are 
given by
\begin{equation}\label{planewave}
 \vec{k}_1 = k(\cos\theta,\sin\theta) \quad\mbox{and}\quad
 \vec{k}_2 = k(\cos\theta,-\sin\theta)\; ,
\end{equation}
where we used $|\vec{k}_1|=|\vec{k}_2|=k$. The wave number $k$ is related to 
the wavelength of the light according to $k=2\pi/\lambda$.

In order to find the interference pattern in the intensity $I$, we sum the 
two plane waves at position $\vec{r}$ at the amplitude level:
\begin{equation}
 I(\vec{r}) \propto \left|e^{i\vec{k}_1\cdot\vec{r}}+e^{i\vec{k}_2\cdot\vec{r}}
 \right|^2 = 4\cos^2\left[ \frac{1}{2}(\vec{k}_1 - \vec{k}_2)\cdot\vec{r} 
 \right]\; .
\end{equation}
When we calculate the inner product $(\vec{k}_1 - \vec{k}_2)\cdot\vec{r}/2$ 
from Eq.\ (\ref{planewave}) we obtain the expression 
\begin{equation}\label{cos2}
 I(x) \propto \cos^2(kx\sin\theta)
\end{equation}
for the intensity along the substrate in direction $x$.

The Rayleigh criterion states that the minimal resolvable feature size 
$\Delta x$ corresponds to the distance between an intensity maximum and an 
adjacent minimum. From Eq.\ (\ref{cos2}) we obtain
\begin{equation}
 k\Delta x\sin\theta = \frac{\pi}{2}\; .
\end{equation} 
This means that the maximum resolution is given by
\begin{equation}
 \Delta x = \frac{\pi}{2k\sin\theta} = \frac{\pi}{2\left(\frac{2\pi}{\lambda}
 \sin\theta\right)} = \frac{\lambda}{4\sin\theta}\; ,
\end{equation}
where $\lambda$ is the wavelength of the light. The maximum resolution is
therefore proportional to the wavelength and inversely proportional to the 
sine of the angle between the incoming plane waves and the normal. The 
resolution is thus maximal ($\Delta x$ is minimal) when $\sin\theta=1$, or 
$\theta=\pi/2$. This is the grazing limit. The classical diffraction limit is 
therefore $\Delta x = \lambda/4$. Note that this derivation does not use the 
approximation $\sin\theta\simeq\theta$, which is common when considering 
diffraction phenomena.

\section{Introduction to Quantum Lithography}\label{intro}

\begin{figure}[h]
  \label{fig1}
  \begin{center}
  \begin{psfrags}
     \psfrag{a}{$a$}
     \psfrag{b}{$b$}
     \psfrag{t}{$\theta$}
     \psfrag{f}{$\varphi$}
     \psfrag{p}{substrate}
     \epsfxsize=8in
     \epsfbox[-100 20 900 70]{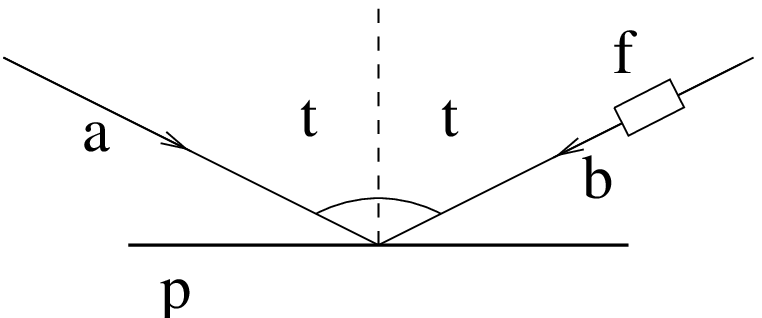}
  \end{psfrags}
  \end{center}
  {\small Fig.\ 1. Two light beams $a$ and $b$ cross each other at the
	surface of a photosensitive substrate. The angle between them is 
	$2\theta$ and they have a relative phase difference $\varphi=kx/2$. We
	consider the limit case of $\theta\rightarrow\pi/2$.}
\end{figure}

In this section we briefly reiterate our method of Ref.\ \cite{boto00}. 
It exploits the physical properties of multi-photon absorption of a substrate.
Suppose we have two intersecting light beams $a$ and $b$. We place a substrate 
sensitive to $N$-photon absorption at the position where the two beams meet, 
such that the interference pattern is recorded. For simplicity, we consider 
the grazing limit in which the angle $\theta$ off axis for the two beams is 
$\pi/2$ (see Fig.\ \ref{fig1}). Classically, the interference pattern on the 
substrate 
has a resolution of the order of $\lambda/4$, where $\lambda$ is the wavelength
of the light. However, by using entangled photon-number states (i.e., 
inherently {\em non-classical} states) we can increase the resolution well 
into the sub-wavelength regime.

How does quantum lithography work? Let the two counter-propagating light beams 
$a$ and $b$ be in the combined entangled state of $N$ photons
\begin{equation}\label{n00n}
 |\psi_N\rangle_{ab} = \left( |N,0\rangle_{ab} + e^{iN\varphi} |0,N\rangle_{ab}
 \right) / \sqrt{2}\; ,
\end{equation}
where $\varphi=kx/2$, with $k=2\pi/\lambda$.
We define the mode operator $\hat{e} = (\hat{a}+\hat{b})/\sqrt{2}$ and its 
adjoint $\hat{e}^{\dagger} = (\hat{a}^{\dagger}+\hat{b}^{\dagger})/\sqrt{2}$. 
The deposition rate $\Delta$ on the substrate is then given by
\begin{equation}\label{delta}
 \Delta_N = \langle\psi_N|\hat{\delta}_N|\psi_N\rangle\qquad\text{with}\qquad
 \hat{\delta}_N=\frac{(\hat{e}^{\dagger})^N \hat{e}^N}{N!}\; ,
\end{equation}
i.e., we look at the higher moments of the electric field operator 
\cite{goppert31,javanainen90,perina98}.
The deposition rate $\Delta$ is measured in units of intensity. Leaving the 
substrate exposed for a time $t$ to the light source will result in an 
exposure pattern $P(\varphi)=\Delta_N t$. After a straightforward calculation 
we see that 
\begin{equation}\label{deprate}
 \Delta_N \propto (1 + \cos N\varphi)\; .
\end{equation}
We interpret this as follows. A path-differential phase-shift $\varphi$ in 
light beam $b$ results in a displacement $x$ of the interference pattern on 
the substrate. Using two classical waves, a phase-shift of $2\pi$ will return 
the pattern to its original position. However, according to 
Eq.~(\ref{deprate}), one cycle is completed after a shift of $2\pi/N$. This 
means that a shift of $2\pi$ will displace the pattern $N$ times. In other 
words, we have $N$ times more maxima in the interference pattern. These need 
to be closely spaced, yielding an effective Rayleigh resolution of $\Delta x
= \lambda/4N$, a factor of $N$ below the classical interferometric result of 
$\Delta x=\lambda/4$ \cite{brueck98}.

Physically, we can interpret this result as follows: instead of having a state
of $N$ single photons, Eq.~(\ref{n00n}) describes an $N$-photon state. Since 
the momentum of this state is $N$ times as large as the momentum for a 
single photon, the corresponding DeBroglie wavelength is $N$ times smaller. 
The interference of this $N$-photon state with itself on a substrate 
thus gives a periodic pattern with a characteristic resolution dimension of 
$\Delta x = \lambda/4N$.

\section{General Patterns in 1D}\label{1D}

So far, we have described a method to print a simple pattern of evenly
spaced lines of sub-wavelength resolution. However, for any practical
application we need the ability to produce more complicated patterns. To this 
end, we introduce the state
\begin{eqnarray}\label{nm}
 |\psi_{Nm}\rangle_{ab} &=& \biggl( e^{im\varphi}|N-m,m\rangle_{ab} \biggr.\cr
 && + \biggl. e^{i(N-m)\varphi} e^{i\theta_m} |m,N-m\rangle_{ab} \biggr) 
 / \sqrt{2}\; .
\end{eqnarray}
This is a generalized version of Eq.\ (\ref{n00n}). In particular, Eq.\
(\ref{nm}) reduces to Eq.\ (\ref{n00n}) when $m=0$ and $\theta_m=0$. Note
that we included a relative phase $e^{i\theta_m}$, which will turn out to be 
crucial in the creation of arbitrary one-dimensional patterns.

We can calculate the deposition rate again according to the procedure in Sec.\ 
\ref{intro}. As we shall see later, in general, we can have superpositions
of the states given by Eq.\ (\ref{nm}). We therefore have to take into account 
the possibility of different values of $m$, yielding a quantity
\begin{equation}
 \Delta_{Nm}^{Nm'} = \langle\psi_{Nm}|\hat{\delta}_N|\psi_{Nm'}\rangle\; .
\end{equation}
Note that this deposition rate depends not only on the parameter $\varphi$,
but also on the relative phases $\theta_m$ and $\theta_{m'}$. The deposition
rate then becomes
\end{multicols}

\noindent\rule{5cm}{.5 pt}

\begin{eqnarray}\label{nmmn}
 \Delta_{Nm}^{Nm'} \propto \sqrt{\binom{N}{m}\binom{N}{m'}} && \left[ 
 e^{i(m'-m)\varphi} + e^{i(N-m-m')\varphi} e^{i\theta_{m'}} + \right. \cr
 && \left. e^{-i(N-m-m')\varphi} e^{-i\theta_m} + e^{-i(m'-m)\varphi} 
 e^{i(\theta_{m'}-\theta_m)} \right] \; ,
\end{eqnarray}

\hfill\noindent\rule{5cm}{.5 pt}

\medskip

\begin{multicols}{2}
\noindent where $\binom{N}{m}$ means $N!/(N-m)! m!$.
Obviously, $\langle\psi_{Nm}|\hat{\delta}_l|\psi_{N'm'}\rangle=0$ when 
$l\not\in \{N,N'\}$. 
For $m=m'$, the deposition rate takes on the form
\begin{equation}\label{genn00n}
 \Delta_{Nm} \propto \binom{N}{m} \left\{ 1 + \cos[(N-2m)\varphi+\theta_m]
 \right\}\; ,
\end{equation}
which, in the case of $m=0$ and $\theta_m=0$, coincides with Eq.\ 
(\ref{deprate}). When $\theta_m$ is suitably chosen, we see that we also have 
access to deposition rates $(1-\cos N\varphi)$ and $(1\pm\sin N\varphi)$.
Apart from this extra phase freedom, Eq.~(\ref{genn00n}) does not look like 
an improvement over Eq.\ (\ref{deprate}), since $N-2m\leq N$, which means that 
the resolution decreases. However, we will show later how these states {\em 
can} be used to produce non-trivial patterns.

First, we look at a few special cases of $\theta_m$ and $\theta_{m'}$. When 
we write $\Delta_{Nm}^{Nm'}=\Delta_{Nm}^{Nm'}(\theta_m,\theta_{m'})$ we have
\begin{mathletters}
\begin{eqnarray}
 \Delta_{nm}^{Nm'} (0,0) &\propto& \cos\Bigl(\frac{N-2m}{2}\,
 \varphi\Bigr) \cos\Bigl(\frac{N-2m'}{2}\,\varphi\Bigr)\; , \\
 \Delta_{Nm}^{Nm'} (0,\pi) &\propto& \cos\Bigl(\frac{N-2m}{2}\,
 \varphi\Bigr) \sin\Bigl(\frac{N-2m'}{2}\,\varphi\Bigr)\; , \\
 \Delta_{Nm}^{Nm'} (\pi,0) &\propto& \sin\Bigl(\frac{N-2m}{2}\,
 \varphi\Bigr) \cos\Bigl(\frac{N-2m'}{2}\,\varphi\Bigr)\; , \\
 \Delta_{Nm}^{Nm'} (\pi,\pi) &\propto& \sin\Bigl(\frac{N-2m}{2}\,
 \varphi\Bigr) \sin\Bigl(\frac{N-2m'}{2}\,\varphi\Bigr)\; .
\end{eqnarray}
\end{mathletters}
These relations give the dependence of the matrix elements $\Delta_{Nm}^{Nm'}$
on $\theta_m$ and $\theta_{m'}$ in a more intuitive way than Eq.\ (\ref{nmmn})
does. Finally, when $\theta_m = \theta_{m'}=\theta$ we obtain
\begin{equation}
 \Delta_{Nm}^{Nm'} \propto \cos\left[\frac{(N-2m)\varphi+\theta}{2}\right]
 \cos\left[\frac{(N-2m')\varphi-\theta}{2}\right]\; .
\end{equation}

So far we have only considered generalized deposition rates given by Eq.\ 
(\ref{nm}), with special values of their parameters. We will now turn our 
attention to the problem of creating more arbitrary patterns.

Note that there are two main, though fundamentally different, ways we can 
superpose the states given by Eq.\ (\ref{nm}). We can superpose states with 
different photon numbers $n$ and a fixed distribution $m$ over the two modes:
\begin{equation}\label{phot}
 |\Psi_m\rangle = \sum_{n=0}^N \alpha_n |\psi_{nm}\rangle\; ,
\end{equation}
with $\alpha_n$ complex coefficients. This is a superposition of states with
{\em different} total photon number in each branch.
Alternatively, we can superpose states with a fixed photon number $N$, but with
different distributions $m$:
\begin{equation}\label{dist}
 |\Psi_N\rangle = \sum_{m=0}^{\lfloor N/2\rfloor} \alpha_m|\psi_{Nm}\rangle\; ,
\end{equation}
where $\lfloor N/2\rfloor$ denotes the largest integer $l$ with $l\leq N/2$ and
$\alpha_m$ again the complex coefficients. Every branch in this superposition
is an $N$-photon state.

These two different superpositions can be used to tailor patterns which are
more complicated than just closely spaced lines. We will now study these two 
different methods.

\subsection{The Pseudo-Fourier Method}

The first method, corresponding to the superposition given by Eq.\ 
(\ref{phot}), we will call the pseudo-Fourier method (this choice of name will 
become clear shortly). When we calculate the deposition rate $\Delta_m$ 
according to the state $|\Psi_m\rangle$ we immediately see that branches with 
different photon numbers $n$ and $n'$ do not exhibit interference:
\begin{equation}
 \Delta_m = \sum_{n=0}^N |\alpha_n|^2 \langle\psi_{nm}|\hat{\delta}_n|
 \psi_{nm}\rangle = \sum_{n=0}^N |\alpha_n|^2 \Delta_{nm} \; .
\end{equation}
Using Eq.\ (\ref{genn00n}) the exposure pattern $P(\varphi)=\Delta_m t$ 
becomes
\begin{equation}\label{fourier}
 P(\varphi) = t \sum_{n=0}^{N} c_n \left\{1+\cos[(n-2m)\varphi+\theta_n] 
 \right\}\; , 
\end{equation}
where $t$ is the exposure time and the $c_n$ are real and positive. Since 
$m<n$ and 
$m$ is fixed, we have $m=0$. We will now prove that this is a Fourier series 
up to a constant. 

A general Fourier expansion of $p(\varphi)$ can be written as
\begin{equation}
 P(\varphi) = \sum_{n=0}^N (a_n \cos n\varphi + b_n \sin n\varphi)\; .
\end{equation}
Writing Eq.\ (\ref{fourier}) as 
\begin{equation}
 P(\varphi) = t\sum_{n=0}^{N} c_n + t \sum_{n=0}^{N} c_n \cos(n\varphi+
 \theta_n)\; ,
\end{equation}
where $t\sum_{n=0}^{N} c_n$ is a constant. If we ignore this constant (its
contribution to the deposition rate will give a general uniform background 
exposure of the substrate, since it is independent of $\varphi$) we see that 
we need
\begin{equation}
 c_n \cos(n\varphi+\theta_n) = a_n \cos n\varphi + b_n \sin n\varphi
\end{equation}
with $c_n$ positive, $\theta_n\in[0,2\pi)$ and $a_n$, $b_n$ real. Expanding 
the left-hand side and equating terms in $\cos n\varphi$ and $\sin n\varphi$
we find
\begin{equation}
 a_n = c_n \cos\theta_n \qquad\mbox{and}\qquad b_n = c_n \sin\theta_n\; .
\end{equation}
This is essentially a co-ordinate change from Cartesian to polar co-ordinates.
Thus, Eq.\ (\ref{fourier}) is equivalent to a Fourier series up to an additive 
constant. Since in the limit of $N\rightarrow\infty$ a Fourier series can 
converge to any well-behaved pattern $P(\varphi)$, this procedure allows us to 
approximate arbitrary patterns in one dimension (up to a constant). It is now 
clear why we call this procedure the pseudo-Fourier method.

However, there is a drawback with this procedure. The deposition rate
$\Delta$ is a positive definite quantity, which means that once the substrate 
is exposed at a particular Fourier component, there is no way this can be
undone. Technically, Eq.\ (\ref{fourier}) can be written as
\begin{equation}
 P(\varphi) = Q \cdot t + t \sum_{n=0}^{N} (a_n\cos n\varphi + 
 b_n\sin n\varphi)\; ,
\end{equation}
where $Q$ is the uniform background `penalty exposure rate' $Q = \sum_{n=0}^{N}
c_n$ we mentioned earlier. The second term on the right-hand side is a true 
Fourier series. Thus in the pseudo-Fourier method there is always a minimum 
exposure of the substrate. Ultimately, this penalty can be traced to the 
absence of interference between the terms with different photon number in Eq.\ 
(\ref{phot}). Next, we will investigate whether our second method of tailoring 
patterns can remove this penalty exposure.

\subsection{The Superposition Method}

We will now study our second method of tailoring patterns, which we call the 
`superposition method' (lacking a better name). Here we keep the total number
of photons $N$ constant, and change how the photons are distributed between
the two beams in each branch [see Eq.\ (\ref{dist})]. A distinct advantage of 
this method is that it {\em does} exhibit interference between the different 
branches in the superposition, which eliminates the uniform background penalty 
exposure. 

Take for instance a superposition of two distinct terms
\begin{equation}
 |\Psi_N\rangle = \alpha_m |\psi_{Nm}\rangle +\alpha_{m'}|\psi_{Nm'}\rangle\; ,
\end{equation}
with $|\alpha_m|^2 + |\alpha_{m'}|^2 =1$ and $|\psi_{nm}\rangle$ given by Eq.\ 
(\ref{nmmn}). After some algebraic manipulation the deposition rate can be
written as
\begin{eqnarray}
 \Delta_N &\propto& |\alpha_m|^2 
 \binom{N}{m} \left\{1+\cos[(N-2m)\varphi + \theta_m] \right\}\cr && 
 + |\alpha_{m'}|^2 \binom{N}{m'} \left\{1+\cos[(N-2m')\varphi + \theta_{m'}]
 \right\}\cr && 
  +8 r_m^{m'}\sqrt{\binom{N}{m}\binom{N}{m'}}
 \cos\left(\frac{\theta_{m'}}{2}-\frac{\theta_m}{2} + \xi_m^{m'} \right)\cr &&
 ~\times \cos\frac{1}{2}\left[(N-2m)\varphi + \theta_m \right] \cr &&
 ~\times \cos\frac{1}{2}\left[(N-2m')\varphi+ \theta_{m'} \right]\; ,
\end{eqnarray}
where the deposition rate $\Delta$ is now a function of 
$\alpha_m$ and $\alpha_{m'}$, where we have chosen the real numbers $r_m^{m'}$ 
and $\xi_m^{m'}$ to satisfy $\alpha_m^*\alpha_{m'} \equiv r_m^{m'}
\exp(i\xi_m^{m'})$. For the special values $N=20$, $m=9$, $m'=5$ and 
$\theta_m=\theta_{m'}=0$ we obtain the pattern shown in Fig.\ \ref{fig3}. 
Clearly, there is no uniform background penalty exposure here. 

\begin{figure}[h]
  \label{fig2a}
  \begin{center}
  \begin{psfrags}
     \psfrag{D}{$\Delta_{20}$, with $m=5$ and $m'=9$.}
     \psfrag{phi}{$\varphi$}
     \psfrag{1.5708}{$\quad\pi/2$}
     \psfrag{Pi}{$~\pi$}
     \psfrag{4.71239}{$\quad 3\pi/2$}
     \psfrag{2 Pi}{$~2\pi$}
     \epsfxsize=8in
     \epsfbox[-20 50 730 240]{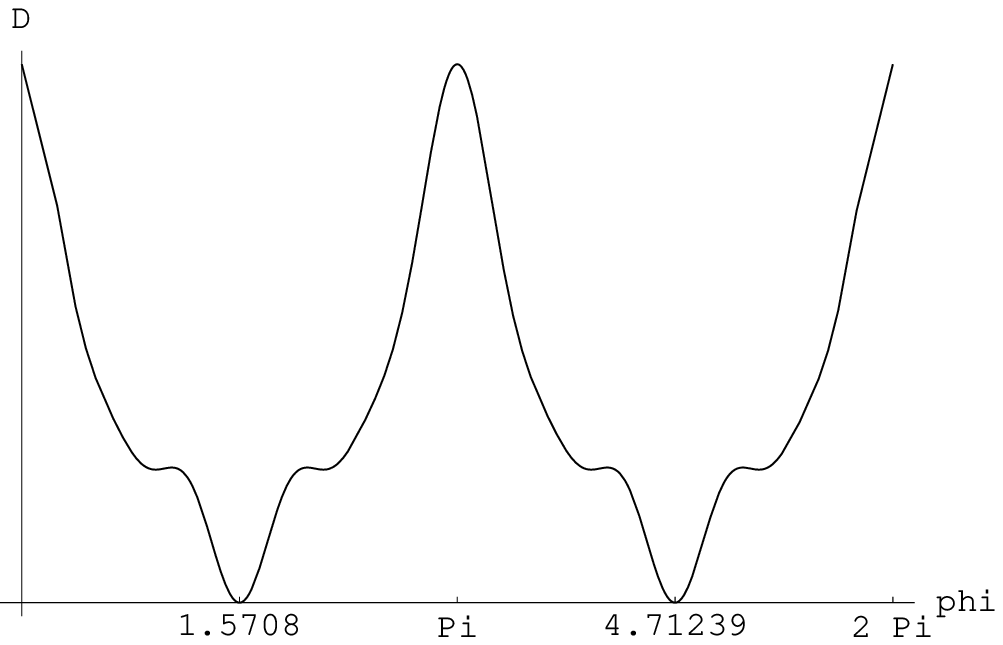}
  \end{psfrags}
  \end{center}
  {\small Fig.\ 2: The deposition rate due to a superposition of two states 
	containing 20 photons with distributions $m=9$ and $m'=5$ ($\theta_m = 
	\theta_{m'}=0$). The deposition rate at $\varphi=\pi/2$ and 
	$\varphi=3\pi/2$ is zero, which means that there is no general uniform 
	background exposure.}
\end{figure}

For more than two branches in the superposition this becomes a  complicated 
function, which is not nearly as well understood as a Fourier series. The 
general expression for the deposition rate can be written as
\begin{eqnarray}
 \Delta_N &\propto& \sum_{m=0}^{\lfloor N/2\rfloor}
 \sum_{m'=0}^{\lfloor N/2\rfloor} r_m^{m'}
 \sqrt{\binom{N}{m}\binom{N}{m'}} \cr
 && \quad~\times \cos\left(\frac{\theta_{m'}}{2}-\frac{\theta_m}{2} +\xi_m^{m'}
    \right)\cr 
 && \quad~\times \cos\frac{1}{2}\left[(N-2m)\varphi + \theta_m\right] \cr
 && \quad~\times\cos\frac{1}{2}\left[(N-2m')\varphi + \theta_{m'} \right]\; ,
\end{eqnarray}
where we have chosen $r_m^{m'}$ and $\xi_m^{m'}$ real to satisfy 
$\alpha_m^*\alpha_{m'} \equiv r_m^{m'}\exp(i\xi_m^{m'})$. Note that 
$\xi_m^m=0$. 

If we want to tailor a pattern $F(\varphi)$, it might be the case that this 
type of superposition will also converge to the required pattern. We will now 
compare the superposition method with the Fourier method.

\subsection{Comparing the two methods}\label{arb}

So far, we discussed two methods of creating non-trivial patterns in one 
dimension. The Fourier method is simple but yields a uniform background 
penalty exposure. The superposition method is far more complicated, but seems 
to get around the background exposure. Before we make a comparison between the 
two methods we will discuss the creation of `arbitrary' patterns.

It is well known that any sufficiently well-behaved periodic function can be 
written as an infinite Fourier series (we ignore such subtleties which arise 
when two functions differ only at a finite number of points, etc.). However, 
when we create patterns with the pseudo-Fourier lithography method we do not 
have access to every component of the Fourier expansion, since this would 
involve an infinite number of photons ($n\rightarrow\infty$). This means that 
we can only employ truncated Fourier series, and these can merely approximate 
arbitrary patterns.

The Fourier expansion has the nice property that when a series is truncated
at $N$, the remaining terms still give the best Fourier expansion of the 
function up to $N$. In other words, the coefficients of a truncated Fourier 
series are equal to the first $N$ coefficients of a full Fourier series. If 
the full Fourier series is denoted by $F$ and the truncated series by $F_N$,
we can define the normed-distance quantity $D_N$:
\begin{equation}
 D_N \equiv \int_0^{2\pi} |F(\varphi)-F_N(\varphi)|^2 d\varphi\; ,
\end{equation}
which can be interpreted as a distance between $F$ and $F_N$. If quantum 
lithography yields a pattern $p_N(\varphi)=\Delta_N t$, we can introduce the 
following definition: quantum lithography can approximate arbitrary patterns 
if 
\begin{equation}
 \int_0^{2\pi} |F(\varphi)-P_N(\varphi)|^2 d\varphi \leq \varepsilon D_N\; ,
\end{equation} 
with $\varepsilon$ some real, positive definite proportionality constant. 
This definition gives the concept of approximating patterns a solid basis.

We compare the Fourier and the superposition method for one special case. We 
choose the test function
\begin{eqnarray}\label{test}
 F(\varphi) = \left\{
 \begin{matrix}
  h ~\mbox{if}~ -\frac{\pi}{2} < \varphi < \frac{\pi}{2}\; , \cr
  \, 0 ~\mbox{otherwise}\; .\quad\qquad
 \end{matrix}
 \right.
\end{eqnarray}
With up to ten photons, we ask how well the Fourier and the superposition 
method approximate this pattern.

In the case of the Fourier method the solution is immediate. The Fourier
expansion of the `trench' function given by Eq.\ (\ref{test}) is well known:
\begin{equation}\label{fourmin}
 F(\varphi) = \sum_{q=0}^{\infty} \frac{(-1)^q}{2q+1} \cos[(2q+1)\varphi]\; .
\end{equation}
Using up to $n=10$ photons we include terms up to $q=4$, since $2q+1\leq 10$.
The Fourier method thus yields a pattern $P(\varphi)$ (the two patterns 
$P(\varphi)$ and $F(\varphi)$ are generally not the same) which can be written 
as 
\begin{equation}
 P(\varphi) = \sum_{q=0}^4 \frac{c_q t}{2q+1}
 \left( 1+\cos\left[(2q+1)\varphi + \pi\kappa_q\right] \right)\; ,
\end{equation}
where $c_q$ is a constant depending on the proportionality constant of 
$\Delta_{2q+1}$, the rate of production of $|\psi_{nm}\rangle$ and the coupling
between the light field and the substrate. The term $\kappa_q$ is defined to
accommodate for the minus signs in Eq.\ (\ref{fourmin}): it is zero when $q$ 
is even and one when $q$ is odd. Note the uniform background penalty exposure 
rate $\sum_{q=0}^4 c_q/(2q+1)$. The result of this method is shown in 
Fig.\ \ref{fig3}.

Alternatively, the superposition method employs a state
\begin{equation}
 |\Psi_{N}\rangle=\sum_{m=0}^{\lfloor N/2\rfloor}\alpha_m|\psi_{Nm}\rangle\; .
\end{equation}
The procedure of finding the best fit with the test function is more 
complicated. We have to minimize the absolute difference between the 
deposition rate $\Delta_{N}(\vec\alpha)$ times the exposure time $t$ and the 
test function $F(\varphi)$. We have chosen $\vec\alpha=(\alpha_0,\ldots,
\alpha_{n/2})$. Mathematically, we have to evaluate the $\vec\alpha$ and $t$ 
which minimize $d_N$:
\begin{equation}\nonumber
 d_N = \int_0^{2\pi} |F(\varphi)-\Delta_N (\vec\alpha) t|^2 d\varphi\; ,
\end{equation}
with
\begin{equation}
 \Delta_N (\vec\alpha) = \langle\Psi_N |\hat{\delta}_N |\Psi_N \rangle\; .
\end{equation}
We have to fit both $t$ and $\vec\alpha$. Using a genetic optimalization 
algorithm \cite{price} (with $h=1$, a normalized height of the test function) 
we found that the deposition rate is actually very close to zero in the 
interval $\pi/2 \leq\varphi\leq 3\pi/2$, unlike the pseudo-Fourier method, 
where we have to pay a uniform background penalty. This result implies that 
in this case a superposition of different photon distributions $m$, given a 
fixed total number of photons $N$, works better than a superposition of 
different photon number states (see Fig.\ \ref{fig3}). In particular, {\em 
the fixed photon number method allows for the substrate to remain virtually 
unexposed in certain areas}.
 
\begin{figure}[h]
  \label{fig2}
  \begin{center}
  \begin{psfrags}
     \psfrag{D}{$P(\varphi) = \Delta_{10}(\varphi)t$}
     \psfrag{phi}{$\varphi$}
     \psfrag{b}{$P(\varphi)\sum_{n=1}^{10} \alpha_n \Delta_n t$}
     \psfrag{pi2}{}
     \psfrag{p}{$\Bigg\updownarrow$ penalty}
     \psfrag{3 pi2}{}
     \psfrag{2 Pi}{$2 \pi$}
     \psfrag{Pi}{$\pi$}
     \psfrag{1.5708}{$\pi/2$}
     \psfrag{4.71239}{$~\quad 3\pi/2$}
     \epsfxsize=8in
     \epsfbox[0 70 700 240]{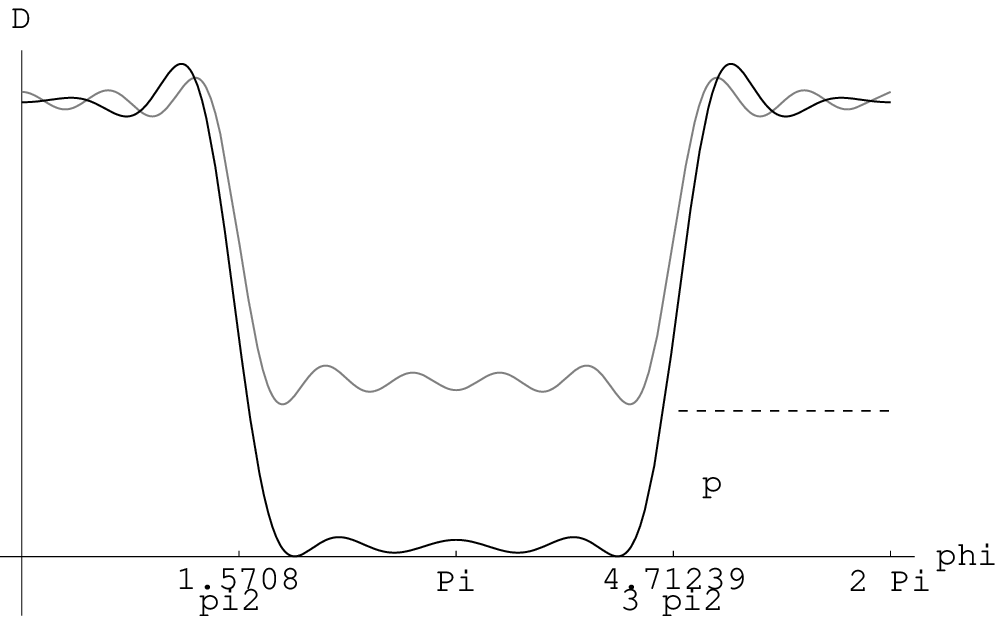}
  \end{psfrags}
  \end{center}
  {\small Fig.\ 3: The deposition rate on the substrate resulting from a 
	superposition of states with $N=10$ and different $m$ (black curve),
	and also resulting from a superposition of states with different $n$ 
	with $m=0$ (grey curve). The coefficients of the superposition that 
	yield the black curve are optimized using a genetic algorithm 
	\cite{price}, while the grey curve is a truncated pseudo-Fourier 
	series. Notice the `penalty' (displacement from zero) of the deposition
	rate for the pseudo-Fourier series between $\pi/2$ and $3\pi/2$.}
\end{figure}

We stress that this is merely a comparison for a specific example, namely that
of the trench target function $F(\varphi)$. We conjecture that the 
superposition method can approximate other arbitrary patterns equally well, 
but we have not yet found a proof. Besides the ability to fit an arbitrary 
pattern, another criterion of comparison between the pseudo-Fourier method and 
the superposition method, is the time needed to create the $N$-photon 
entangled states. 

Until now, we have only considered sub-wavelength resolution in one direction,
namely parallel to the direction of the beams. However, for practical
applications we would like sub-wavelength resolution in both directions on
the substrate. This is the subject of the next section.

\section{General Patterns in 2D}\label{2D}

In this section we study how to create two-dimensional patterns
on a suitable substrate using the quantum lithography techniques developed
in the previous sections. As we have seen, the phase shift $\varphi$, in the
setup given by Fig.\ \ref{fig1}, acts as a parametrization for the deposition 
rate in one dimension. Let's call this the $x$-direction.

\begin{figure}[h]
  \label{fig3}
  \begin{center}
  \begin{psfrags}
     \psfrag{a}{$a$}
     \psfrag{b}{$b$}
     \psfrag{c}{$c$}
     \psfrag{d}{$d$}
     \psfrag{t}{$\theta$}
     \psfrag{x}{$\chi$}
     \psfrag{f}{$\varphi$}
     \psfrag{s}{substrate}
     \epsfxsize=8in
     \epsfbox[-100 20 900 220]{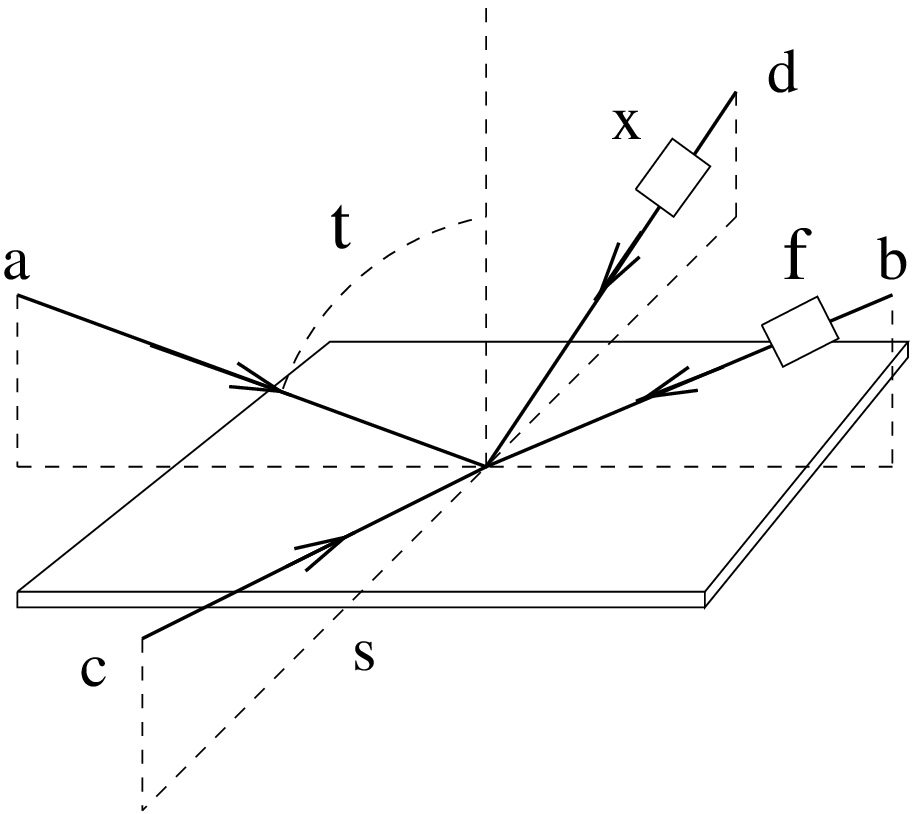}
  \end{psfrags}
  \end{center}
  {\small Fig.\ 4: Four light beams $a$, $b$, $c$ and $d$ cross each other at 
	the surface of a photosensitive substrate. The angles between $a$ and
	$b$ and $c$ and $d$ are again taken in the grazing limit of 
	$\theta=\pi/2$. The relative phase difference between $a$ and $b$ is 
	$\varphi$ and the relative phase difference between $c$ and $d$ is 
	$\chi$.}
\end{figure}

\begin{figure}[h]
  \label{fig4}
  \begin{center}
  \begin{psfrags}
     \psfrag{X}[b]{$y$}
     \psfrag{Y}[b]{$x$}
     \psfrag{Z}{$\Delta$}
     \psfrag{0}{}
     \psfrag{6}{}
     \psfrag{2}{}
     \psfrag{4}{}
     \psfrag{10}{}
     \psfrag{5}{}
     \psfrag{2.5}{}
     \psfrag{7.5}{}
     \epsfxsize=8in
     \epsfbox[0 50 700 250]{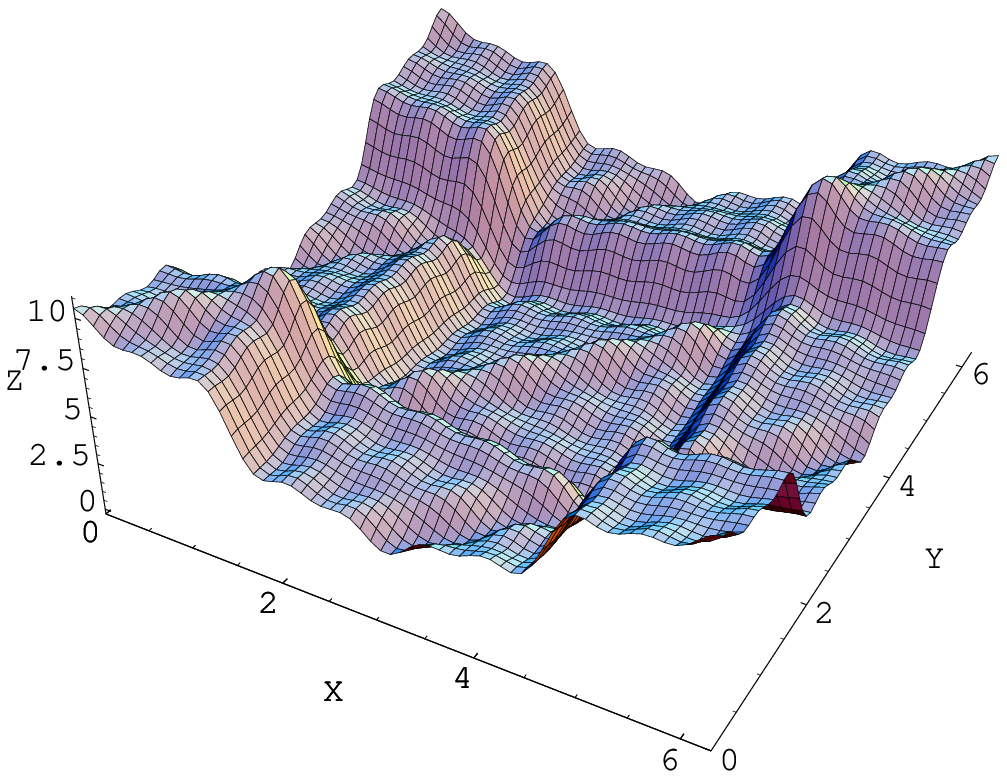}
  \end{psfrags}
  \end{center}
  {\small Fig.\ 5: A simulation of a two-dimensional intensity pattern on an 
	area $\lambda^2$, where $\lambda$ denotes the wavelength of the used 
	light. Here we modelled a square area with sharp edges. The pattern 
	was generated by a Fourier series of up to ten photons (see also
	Fig.\ 3 for the one-dimensional case).}
\end{figure}

We can now do the same for the $y$-direction, employing two counter-propagating
beams ($c$ and $d$) in the $y$-direction (see Fig.\ \ref{fig4}). The same 
conditions apply: we consider the limit where the spatial angle $\theta$ off 
axis approaches $\pi/2$, thus grazing along the substrate's surface. 

Consider the region where the four beams $a$, $b$, $c$ and $d$ overlap. For 
real lithography we have to take into account the mode shapes, but when we 
confine ourselves to an area with side lengths $\lambda$ (where $\lambda$ is 
the wavelength of the used light) this problem does not arise.

The class of states on modes $a$ to $d$ that we consider here are of the form
\begin{eqnarray}\label{nmk}
 |\psi^k_{Nm}\rangle &=& \frac{1}{2} 
 \biggl[ e^{im\varphi}|N-m,m;0,0\rangle \biggr. \cr && 
 \qquad +\; e^{i(N-m)\varphi} e^{i\zeta_m}|m,N-m;0,0\rangle \cr && 
 \qquad +\; e^{ik\theta}|0,0;N-k,k\rangle \cr && 
 \qquad \biggl. +\; e^{i(N-k)\chi} e^{i\bar\zeta_k}|0,0;k,N-k\rangle \biggr]
 \; ,
\end{eqnarray}
where $\zeta_m$ and $\bar\zeta_k$ are two relative phases. 
This is by no means the only class of states, but we will restrict our 
discussion to this one for now. Observe that this is a superposition on the 
amplitude level, which allows destructive interference in the deposition rate 
in order to create dark spots on the substrate. Alternatively, we could have 
used the one-dimensional method [with states given by Eq.\ (\ref{nm})] in the 
$x$- and $y$-direction, but this cannot give interference effects between the 
modes $a,b$ and $c,d$.

The phase-shifts $\varphi$ and $\chi$ in the light beams $b$ and $d$ (see
Fig.\ \ref{fig4}) result in respective displacements $x$ and $y$ of the 
interference 
pattern on the substrate. A phase-shift of $2\pi$ in a given direction will 
displace the pattern, say, $N$ times. This means that the maxima are closer
together, yielding an effective resolution equal to $\Delta x = \Delta y =
\lambda/4N$. This happens in both the $x$- and the $y$-direction. 

We proceed again as in Sec.\ \ref{intro} by evaluating the $N^{\rm th}$ order
moment $\hat{\delta}_N$ of the electric field operator [see Eq.~(\ref{delta})].
This gives the deposition rate $\Delta_{Nmk}^{Nm'k'} = \langle \psi^k_{Nm}|
\delta_N|\psi^{k'}_{Nm'}\rangle$ [with $|\psi^k_{Nm}\rangle$ given by Eq.\ 
(\ref{nmk})]:
\end{multicols}

\noindent\rule{5cm}{.5 pt}

\medskip

\begin{eqnarray}\label{2ddeprate}
 \Delta_{Nmk}^{Nm'k'} \propto &&
    \binom{N}{m}\binom{N}{m'} 
 \left( e^{-im\varphi} e^{im'\varphi} + 
  e^{-im\varphi} e^{i(N-m')\varphi} e^{i\zeta_{m'}} + e^{-i(N-m)\varphi}
  e^{im'\varphi} e^{-i\zeta_m} \right.\cr &&\qquad\qquad\qquad\qquad\left. + 
  e^{-i(N-m)\varphi} e^{i(N-m')\varphi}
  e^{-i(\zeta_m-\zeta_{m'})} \right) \cr
 && + \binom{N}{m}\binom{N}{k'} 
 \left( e^{-im\varphi} e^{ik'\chi} + 
  e^{-im\varphi} e^{i(N-k')\chi} e^{i\bar\zeta_{k'}} + e^{-i(N-m)\varphi}
  e^{ik'\chi} e^{-i\zeta_m} \right.\cr &&\qquad\qquad\qquad\qquad\left. + 
  e^{-i(N-m)\varphi} e^{i(N-k')\chi}
  e^{-i(\zeta_m-\bar\zeta_{k'})} \right) \cr
 && + \binom{N}{k}\binom{N}{m'} 
 \left( e^{-ik\chi} e^{im'\varphi} + 
  e^{-ik\chi} e^{i(N-m')\varphi} e^{i\zeta_{m'}} + e^{-i(N-k)\chi}
  e^{im'\varphi} e^{-i\bar\zeta_k} \right.\cr &&\qquad\qquad\qquad\qquad\left. 
  + e^{-i(N-k)\chi} e^{i(N-m')\varphi}
  e^{-i(\bar\zeta_k-\zeta_{m'})} \right) \cr
 && + \binom{N}{k}\binom{N}{k'} 
 \left( e^{-ik\chi} e^{ik'\chi} + 
  e^{-ik\chi} e^{i(N-k')\chi} e^{i\bar\zeta_{k'}} + e^{-i(N-k)\chi}
  e^{ik'\chi} e^{-i\bar\zeta_k} \right.\cr &&\qquad\qquad\qquad\qquad\left. + 
  e^{-i(N-k)\chi} e^{i(N-k')\chi}
  e^{-i(\bar\zeta_k-\bar\zeta_{k'})} \right)\; .
\end{eqnarray}
For the special choice of $m'=m$ and $k'=k$ we have 
\begin{eqnarray}
 \Delta_{Nm}^{k} &\propto&
 \binom{N}{m}^2 \left( 1 + \cos[(N-2m)\varphi+\zeta_m] \right) 
 + \binom{N}{k}^2 \left( 1 + \cos[(N-2k)\chi+\bar\zeta_k] \right)\cr && 
 + 4 \binom{N}{m} \binom{N}{k} \cos\frac{1}{2}\left[ N(\varphi-\chi)
 + (\zeta_m-\bar\zeta_k) \right] \cr && \qquad\times
 \cos\frac{1}{2}\left[ (N-2m)\varphi-\zeta_m\right] 
 \cos\frac{1}{2}\left[ (N-2k)\chi-\bar\zeta_k\right] \; .
\end{eqnarray}

\hfill\noindent\rule{5cm}{.5 pt}

\medskip

\begin{multicols}{2}
We can again generalize this method and use superpositions of the states 
given in Eq.\ (\ref{nmk}). Note that there are now three numbers $N$, $m$ and 
$k$ which can be varied. Furthermore, as we have seen in the one-dimensional 
case, superpositions of different $n$ do not give interference terms in the 
deposition rate. 

Suppose we want to approximate a pattern $F(\varphi,\chi)$, with $\{\varphi,
\chi\}\in [0,2\pi]$. This pattern can always be written in a Fourier expansion:
\begin{eqnarray}\label{2Dfourier}
 F(\varphi,\chi) &=& \sum_{p,q=0}^{\infty} a_{pq} \cos p\varphi\cos q\chi +
 b_{pq} \cos p\varphi\sin q\chi \times\cr
 && \qquad c_{pq} \sin p\varphi\cos q\chi + d_{pq} \sin p\varphi\sin 
 q\chi \; .
\end{eqnarray}
with $a_{pq}$, $b_{pq}$, $c_{pq}$ and $d_{pq}$ real. In the previous section 
we showed that quantum lithography could approximate the Fourier series of a 
one-dimensional pattern up to a constant displacement. This relied on absence 
of interference between the terms with different photon numbers. The question 
is now whether we can do the same for patterns in {\em two} dimensions. 
Or alternatively, can general superpositions of the state 
$|\psi_{Nm}^k\rangle$ approximate the pattern $F(\varphi,\chi)$?

From Eq.\ (\ref{2ddeprate}) it is not obvious that we can obtain the four 
trigonometric terms given by the Fourier expansion of Eq.\ (\ref{2Dfourier}):
\begin{mathletters}
\begin{eqnarray}
 \Delta &~\propto~& \cos p\varphi\, \cos q\chi\; , \\
 \Delta &~\propto~& \cos p\varphi\, \sin q\chi\; , \\
 \Delta &~\propto~& \sin p\varphi\, \cos q\chi\; , \\
 \Delta &~\propto~& \sin p\varphi\, \sin q\chi\; .
\end{eqnarray}
\end{mathletters}
We can therefore not claim that two-dimensional quantum lithography can 
approximate arbitrary patterns in the sense of one-dimensional lithography.
Only simple patterns like the one given in Fig.\ \ref{fig4} can be inferred 
from Eq.\
(\ref{2ddeprate}). In order to find the best fit to an arbitrary pattern 
one has to use a minimization procedure. 

For example, we calculate the total deposition rate due to the quantum state 
$|\Psi_N\rangle$, where
\begin{equation}
 |\Psi_N\rangle = \sum_{m=0}^{\lfloor N/2\rfloor} \sum_{k=0}^{\lfloor N/2
 \rfloor} \alpha_{mk} |\psi_{Nm}^k\rangle\; .
\end{equation}
Here, $\alpha_{mk}$ are complex coefficients. We now proceed by choosing a 
particular intensity pattern $F(\varphi,\chi)$ and optimising the 
coefficients $\alpha_{mk}$ for a chosen number of photons. The deposition rate 
due to the state $|\Psi_N\rangle$ is now 
\begin{equation}
 \Delta_N (\vec\alpha) = \sum_{m,m'=0}^{\lfloor N/2\rfloor} 
 \sum_{k,k'=0}^{\lfloor N/2\rfloor} \alpha^*_{mk} \alpha_{m'k'} 
 \Delta_{Nmk}^{Nm'k'}\; ,
\end{equation}
with $\vec\alpha=(\alpha_{0,0},\alpha_{0,1}\ldots,\alpha_{N/2,N/2})$.
We again have to evaluate the $\vec\alpha$ and $t$ which minimize
\begin{equation}
 \int_{0}^{2\pi} \int_{0}^{2\pi} \left| F(\varphi,\chi)
 - \Delta_N (\vec\alpha) t \right|^2 d\varphi\, d\chi\; .
\end{equation}
The values of $\vec\alpha$ and $t$ can again be found using a genetic 
algorithm.

\section{Physical implementation}\label{phys}

With current experimental capabilities, the physical implementation of quantum 
lithography is very challenging. In particular, there are two major issues to
be dealt with before quantum lithography can become a mature technology. First
of all, we not only need the ability to create the entangled photon states 
given by Eqs.\ (\ref{nm}) and (\ref{nmk}), but we should also be able to 
create coherent superpositions of these states. One possibility might be to 
use optical components like parametric down-converters. Contrary to the 
results of Ref.\ \cite{kok}, we are not concerned with the usually large 
vacuum contribution of these processes, since the vacuum will not contribute 
to the spatial profile of the deposition [see Eqs.\ (\ref{n00n}) and 
(\ref{delta})]. 

Secondly, we need substrates which are sensitive to the higher moments of the 
electric field operator. When we want to use the pseudo-Fourier method, up to 
$N$ photons for quantum lithography in one dimension, the substrate needs to be
reasonably sensitive to all the higher moments up to $N$, the maximum photon 
number. Alternatively, we can use the superposition method for $N$ photons 
when the substrate is sensitive to predominantly one higher moment 
corresponding to $N$ photons. Generally, the method of lithography determines 
the requirements of the substrate.

There are also some considerations about the approximation of patterns. For
example, we might not {\em need} arbitrary patterns. It might be the case that 
it is sufficient to have a set of patterns which can then be used to generate 
any desired circuit. This is analogous to having a universal set of logical 
gates, permitting any conceivable logical expression. In that case we only 
need to determine this elementary set of patterns.

Furthermore, we have to study whether the uniform background penalty exposure 
really presents a practical problem. One might argue that a sufficient 
difference between the maximum deposition rate and the uniform background 
penalty exposure is enough to accommodate lithography. This depends on 
the details of the substrate's reaction to the electro magnetic field.

Before quantum lithography can be physically implemented and used in the 
production of nano circuits, these issues have to be addressed satisfactorily.

\section{Conclusions}

In this paper we have generalized the theory of quantum lithography as first 
outlined in Ref.\ \cite{boto00}. In particular, we have shown how we can create
arbitrary patterns in one dimension, albeit with a uniform background
penalty exposure. We can also create some patterns in two dimensions, but 
we have no proof that this method can be extended to give arbitrary patterns.

For lithography in one dimension we distinguish two methods: the 
pseudo-Fourier method' and the superposition method. The pseudo-Fourier method 
is conceptually easier since it depends on Fourier 
analysis, but it also involves a finite amount of unwanted exposure of 
the substrate. More specifically, the deposition rate equals the pattern in
its Fourier basis plus a term yielding unwanted background exposure.
The superposition method gets around this problem and seems to give better 
results, but lacks the intuitive clarity of the Fourier method. Furthermore,
we do not have a proof that this method can approximate arbitrary patterns
(see Sec.\ \ref{arb} for a discussion on this approximation).

Quantum lithography in two dimensions is more involved. Starting with a 
superposition of states, given by Eq.\ (\ref{nmk}), we found that we can indeed
create two-dimensional patterns with sub-wavelength resolution, but we do not
have a proof that we can create {\em arbitrary} patterns. Nevertheless, we 
might be able to create a certain set of elementary basis patterns.

There are several issues to be addressed in the future. First, we 
need to study the specific restrictions on the substrate and how we can 
physically realize them. Secondly, we need to create the various entangled 
states involved in the quantum lithography protocol.

Finally, G.S.\ Agarwal and R.\ Boyd have called to our attention that quantum 
lithography works also if the weak parametric downconverter source, described
in Ref.\ \cite{boto00} is replaced by a high-flux optical parametric 
amplifier \cite{agarwal00}. The visibility saturates at 20\% in the limit of 
large gain, but this is quite sufficient for some lithography purposes, as well
as for 3D optical holography used for data storage.

\section*{acknowledgements}

We would like to acknowledge interesting and useful discussions with G.S.\
Agarwal, R.\ Boyd, D.\ Branning, M.\ Holland, P.G.\ Kwiat, Y.\ Shih, J.E.\ 
Sipe, D.\ Strekalov, R.B.\ Vrijen and E.\ Yablonovich. A portion of the 
research in this paper was carried out at the Jet Propulsion Laboratory,
California Institute of Technology, under a contract with the National 
Aeronautics and Space Administration. In addition, this research was supported
by the Office of Naval Research and by project QUICOV under the IST-FET-QIPC 
programme.

\end{multicols}

\end{document}